\documentclass[conference]{IEEEtran}

\usepackage{subcaption}
\usepackage{hyperref}
\usepackage[table,xcdraw]{xcolor}
\usepackage{makecell}
\usepackage{lipsum}

\usepackage{cite}
\usepackage{amsmath,amssymb,amsfonts}
\usepackage{algorithmic}
\usepackage{graphicx}
\usepackage{textcomp}
\usepackage{xcolor}

\newcommand{\ket}[1]{|#1\rangle}

\begin{document}

\title{Quantum State Fidelity for Functional Neural Network Construction}

\author{\IEEEauthorblockN{Skylar Chan\IEEEauthorrefmark{1} \quad\quad Wilson Smith\IEEEauthorrefmark{2} \quad\quad Kyla Gabriel\IEEEauthorrefmark{3}}

\IEEEauthorblockA{\IEEEauthorrefmark{1}University of Maryland School of Medicine, skylar.chan@som.umaryland.edu\\}

\IEEEauthorblockA{\IEEEauthorrefmark{2}University of Maryland College Park, smith@umd.edu\\}

\IEEEauthorblockA{\IEEEauthorrefmark{3}Harvard Medical School, kyla\_gabriel@hms.harvard.edu\\}

}

\maketitle

\begin{abstract}

Neuroscientists face challenges in analyzing high-dimensional neural recording data of dense functional networks. Without ground-truth reference data, finding the best algorithm for recovering neurologically relevant networks remains an open question. We implemented hybrid quantum algorithms to construct functional networks and compared them with the results of documented classical techniques. We demonstrated that our quantum state fidelity methods can provide competitive alternatives to classical metrics by revealing distinct functional networks. Our results suggest that quantum computing offers a viable and potentially advantageous alternative for data-driven modeling in neuroscience, underscoring its broader applicability in high-dimensional graph inference and complex system analysis.

\end{abstract}


\section{Introduction}

Functional neural networks (FNNs) are powerful tools for identifying and representing how populations of neurons respond to stimuli and interact across space and time. They are necessary for brain-computer interface and drug development. In computational neuroscience, these networks are often inferred from high-dimensional brain recordings such as two-photon calcium imaging or electro-physiological data \cite{grienberger2022two}.

Classical approaches to constructing FNNs typically rely on similarity metrics such as Pearson correlation and Euclidean distance to quantify relationships between neuron response patterns. A full neural network may consist of all-to-all pairwise connections between neurons, but that is too dense to identify meaningful topologies. Therefore, to assess the most important functional relationships, smaller network structures are determined by thresholding, clustering, or graph-theoretical techniques such as minimum spanning trees (MST) and further analyzed for neuroscientific properties \cite{nelson2021neuronal}.

While these classical methods have enabled significant insight into brain organization, one fundamental limitation is that there are often no ground-truth data on which structures are the most meaningful when analyzing novel brain recordings. Different similarity metrics can create vastly different network topologies, so a variety of different metrics may need to be assessed to identify relevant networks \cite{mohanty2020rethinking}. Therefore, identifying new similarity metrics could help accelerate discovery of novel functional networks.

One potential area where novel similarity metrics can be derived is quantum computing (QC), which has been proposed to benefit neuroscience with the ability to embed and process data in high-dimensional Hilbert spaces \cite{emani2021quantum}. A well-documented similarity metric in QC is quantum state fidelity, which has applied to the development of classifiers for facial emotion detection \cite{singh2022emotion} and quantum support vector machines \cite{havlivcek2019supervised}.

\begin{figure*}[ht]
    \centering
    \includegraphics[width=\linewidth]{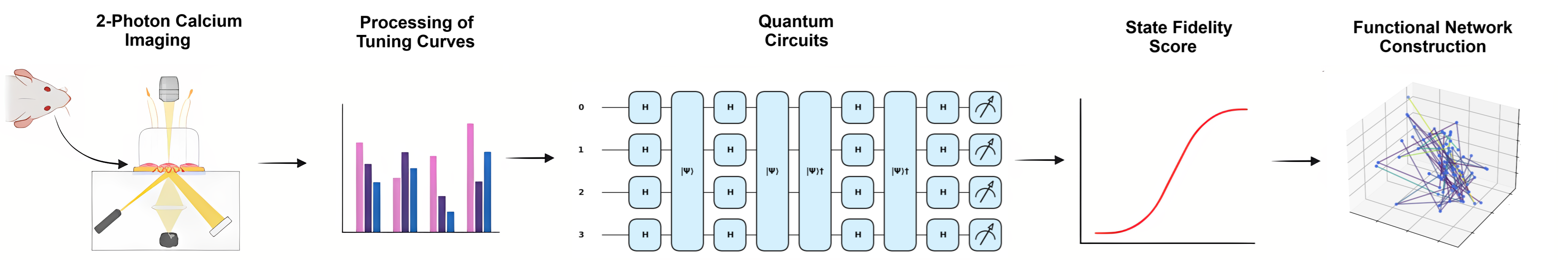}
    \caption{Overview of our method from 2P-imaging and to functional network construction. Figure created with BioRender.}
    \label{fig:main-figure}
\end{figure*}

In this work, we developed a pipeline to compute the quantum state fidelities between embedded pairs of neural tuning curves. The data was sourced from calcium imaging of neuronal responses to auditory stimuli in mice \cite{bowen2024fractured}. We evaluated three quantum embedding methods on quantum simulators and IBM hardware against three classical baseline metrics. Our methods identified new functional networks with statistically significant differences in connectivity patterns versus those extracted using classical techniques.
Within  current scientific literature, our approach, shown in Fig. \ref{fig:main-figure}, is the first application of quantum state fidelity used to uncover functional connectivity of the brain.

\section{Methods}

\subsection{2-Photon Calcium Imaging Dataset}

We used data from a recent 2-photon imaging dataset with recordings of how the auditory cortex in mice responds to pure pitch tones collected from Bowen et al. \cite{bowen2024fractured}. Data was obtained by surgically placing cranial windows on the auditory cortex of the mice and performing 2-photon calcium imaging to give a three-dimensional view of neuron activity in real time. Each neuron's activity is converted into a tuning curve by measuring its normalized fluorescence response  ($\Delta F/F_0$) to nine pure tone stimuli ranging from 3 to 48 kHz, spaced at half-octave intervals. We omit the first and last plane of the smallest recording and extract the neurons with a significant response ($p < 0.01$) to any frequency to obtain 76 neurons positioned in 3D space in the recording plane.

\subsection{State Preparation of Neural Tuning Curves}

We developed three methods to prepare neuron tuning curves as quantum states:

\begin{enumerate}
    \item \textit{Rescale + Angle Embedding} (Ang): The tuning curve is divided by its L1 norm and scaled by $\pi$. The state is uploaded as one layer of Pauli-X rotation gates $R_X(\theta)$, where each rotation is the corresponding value on the tuning curve.
    \item \textit{Resample + Amplitude Embedding} (Amp): Each tuning curve is resampled to have the next power of 2 data points ($9<2^4\rightarrow16=2^4$), preserving endpoints. The modified Akima interpolation method \cite{akima1970new} is used for resampling to create a smooth curve that passes through the original points. Resampled curves are normalized with L2 normalization, and then uploaded with amplitude embedding.
    \item \textit{Resample + Amplitude Embedding \& Fourier Transform} (Amp+QFT): Amplitude embedding with resampling is performed as described above, followed by the Quantum Fourier Transform (QFT) which modifies the amplitudes of quantum states.
\end{enumerate}

\subsection{Quantum State Fidelity Circuits}

We implemented two existing quantum algorithms to compute state fidelity:

\begin{enumerate}
    \item \textit{Hardware-efficient Swap test}: Two states are uploaded onto separate sets of qubits, and each qubit of the first state is entangled with the corresponding qubit of the second state. Hadamard gates are then applied to the first state. All qubits are measured, and the average of the boolean function $f(s,t) = (-1)^{\left( \sum_{i=1}^n s_i t_i \right)}$ over the sampled probabilities is taken as the state fidelity \cite{cincio2018learning}.
    \item \textit{Compute-Uncompute}\: The first state is loaded twice, and the second state is unloaded twice. Loading is prefixed and unloading is postfixed with a layer of Hadamards. Then all qubits are measured, and the probability of $\ket 0$ is taken as the state fidelity \cite{havlivcek2019supervised}.
\end{enumerate}

\section{Experiments}

\subsection{Quantum Metrics}

We tested our three combinations of neural tuning curve embeddings and state fidelity circuits: 1) ``Ang'', 2) ``Amp'', and 3) ``Amp+QFT''.

\subsection{Obtaining State Fidelities}

\subsubsection{Pennylane Simulations}

We implemented Ang, Amp, and Amp+QFT circuits with ideal, noise-free Pennylane Lightning simulators \cite{asadi2024hybrid} and the Pennylane-Catalyst just-in-time compiler \cite{ittah2024catalyst} on one AMD 9800X3D CPU and nVidia RTX 3090 GPU. To reduce simulation time, we used the property that quantum state fidelity is symmetric on noiseless hardware, and only computed state fidelity for each pair of neuron tuning curves $a,b$ where $a < b$. Table \ref{tab:clifford-gate-counts} shows the number of gates per circuit. 

\subsubsection{IBM Hardware}

We used the qBraid SDK (\href{https://docs.qbraid.com/sdk/}{https://docs.qbraid.com/sdk/}) to transpile the Pennylane circuits to Qiskit. We then ran them on the IBM Kingston quantum processing unit (QPU) with 1,000 shots each, and default settings. IBM's backend transpiled our circuits into their native gates. As in the simulation, we only ran circuits to compute state fidelity between pairs of neuron tuning curves $a,b$ where $a < b$. The result of these algorithms on the hardware is noted with ``+IBM.''

\subsubsection{Error Mitigation}

To mitigate potential errors from generic hardware noise, we used the Mitiq toolkit \cite{larose2022mitiq} to apply Digital Dynamic Decoupling (DDD) with XYXY pulses to our amplitude embedding circuits.
This results in ``Amp+DDD'' and ``Amp+QFT+DDD'' circuits, which include additional X and Y gates. DDD was not applied to ``Ang'' circuits as they are too short to insert XYXY pulse trains. Table \ref{tab:ibm-gate-counts} shows the native gate counts of circuits run on the IBM Kingston QPU. The number of native gates tended to increase with DDD mitigation, but in some cases the number of gates decreased after transpilation.

\begin{table*}[!t]

\begin{subtable}[!t]{0.4\textwidth}
\centering
\resizebox{\textwidth}{!}{%
\begin{tabular}{|c|c|c|c|}
\hline
 \textbf{Gates} &
   \textbf{Ang} &
   \textbf{Amp} &
   \textbf{Amp+QFT} \\ \hline
 Num. wires &
   18 &
   4 &
   4 \\ \hline
 Num. gates &
   36 &
   132 &
   156 \\ \hline
 2-qubit gates &
   9 &
   56 &
   72 \\ \hline
 Depth &
   3 &
   108 &
   124 \\ \hline
 Gate types &
 \makecell{
 Hadamard: 9 \\
 RX: 18 \\
 CNOT: 9
 } &
 \makecell{
 Hadamard: 16 \\
RY: 60 \\
CNOT: 56 \\
} &
\makecell{
Hadamard: 24 \\
RY: 60 \\
CNOT: 56 \\
CZ: 12 \\
SWAP: 4 \\
} \\ \hline
\end{tabular}%
}
\caption{Gate counts in our implementation as Clifford gates}
\label{tab:clifford-gate-counts}
\end{subtable}
\hfill
\begin{subtable}[t!]{0.56\textwidth}
\centering
\resizebox{\textwidth}{!}{%
\begin{tabular}{|c|c|c|c|c|c|}
\hline
 \textbf{(IBM)} &
   \textbf{Ang} &
   \textbf{Amp} &
   \textbf{Amp+DDD} &
   \textbf{Amp+QFT} &
   \textbf{Amp+QFT+DDD} \\ \hline
 Num. wires &
   18 &
   4 &
   4 &
   4 &
   4 \\ \hline
 Num. gates &
   126 &
   306-513 &
   334-527 &
   485-697 &
   493-717 \\ \hline
 Depth &
   8 &
   221-325 &
   246-333 &
   242-447 &
   338-447 \\ \hline
 2-qubit gates &
   9 &
   56-104 &
   62-104 &
   94-142 &
   96-140 \\ \hline
 SX gates &
   45 &
   135-236 &
   142-238 &
   222-326 &
   215-323 \\ \hline
 RZ gates &
   54 &
   104-166 &
   115-176 &
   155-224 &
   158-235 \\ \hline
 CZ gates &
   9 &
   56-104 &
   62-104 &
   94-142 &
   96-140 \\ \hline
 X gates &
   0 &
   3-10 &
   3-13 &
   4-9 &
   12-25 \\ \hline
\end{tabular}%
}

\caption{Gate counts after IBM's backend transpilation for Kingston QPU} %
\label{tab:ibm-gate-counts}
\end{subtable}
\caption{Gate counts for each circuit}
\label{tab:gate-count-table}
\end{table*}

\subsection{Classical Metrics}

To provide baseline comparisons for our quantum metrics, we computed three relevant classical metrics for each pair of tuning curves: 1) Pearson correlation (``Correlation'') between tuning curves, 2) Euclidean distance (``Euclidean'') between rescaled tuning curves, and 3) classical fidelity (``Fidelity'') between rescaled tuning curves.

\section{Results}

\subsection{Visual comparison of distance matrices}

\begin{figure}[b!]
    \centering
    \begin{subfigure}{0.4\textwidth}
        \centering
        \includegraphics[width=1\textwidth]{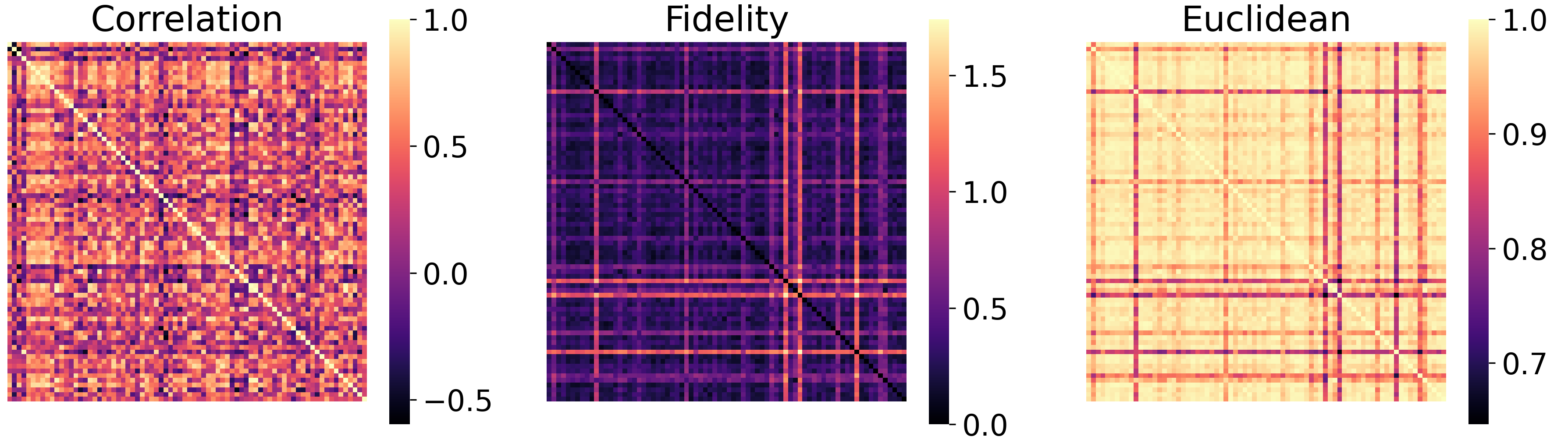}
        \caption{Classical metrics}
        \label{fig:heatmap-ang}
    \end{subfigure}
    \begin{subfigure}{0.28\textwidth}
        \centering
        \includegraphics[width=1\textwidth]{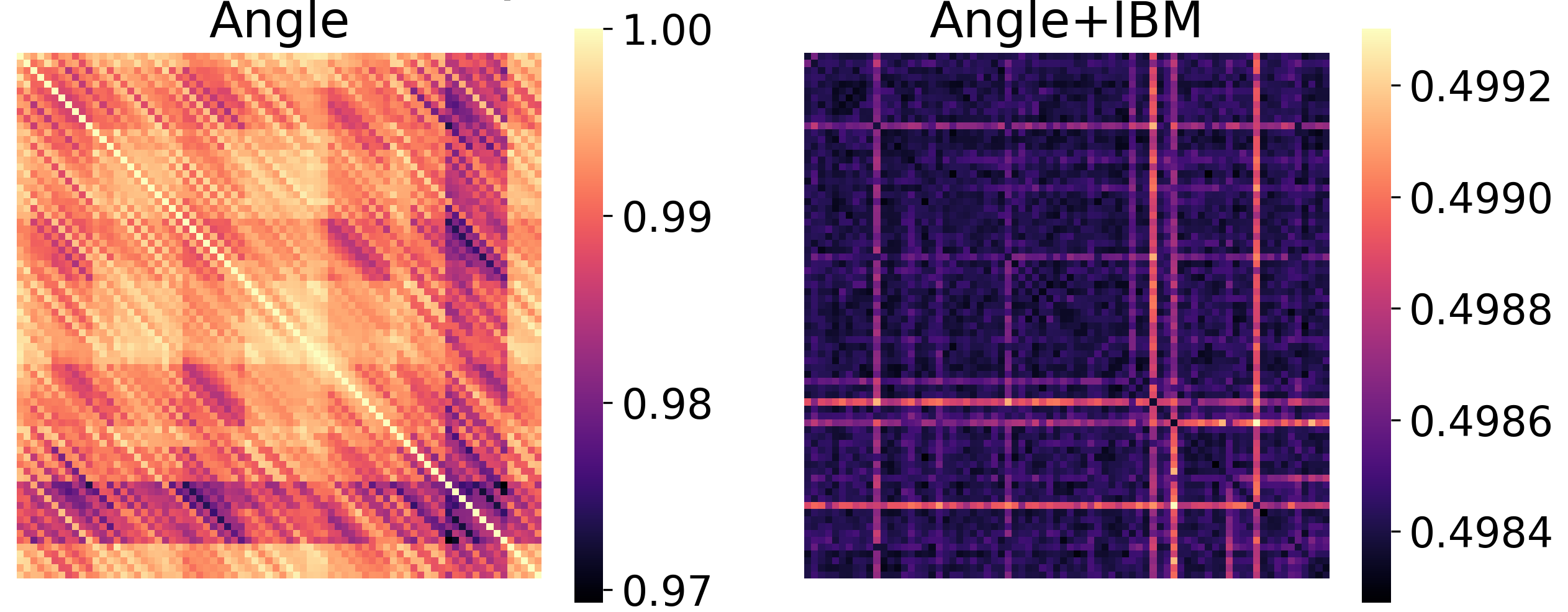}
        \caption{Angle embedding}
        \label{fig:heatmap-ang}
    \end{subfigure}
    \begin{subfigure}{0.4\textwidth}
        \centering
        \includegraphics[width=1\textwidth]{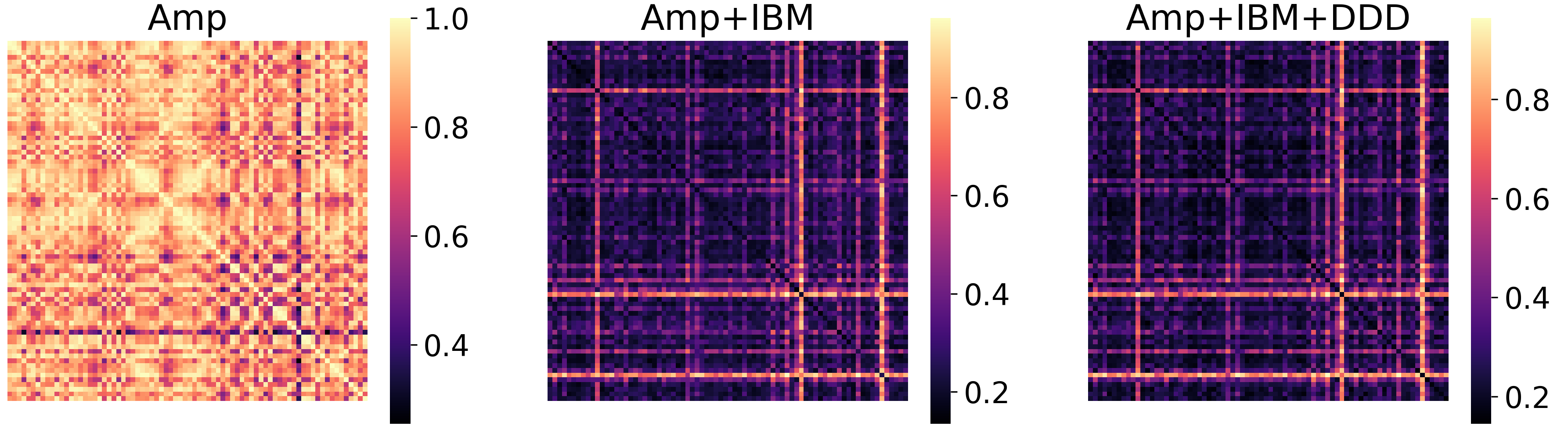}
        \caption{Amplitude embedding}
        \label{fig:heatmap-amp}
    \end{subfigure}
    \begin{subfigure}{0.4\textwidth}
        \centering
        \includegraphics[width=1\textwidth]{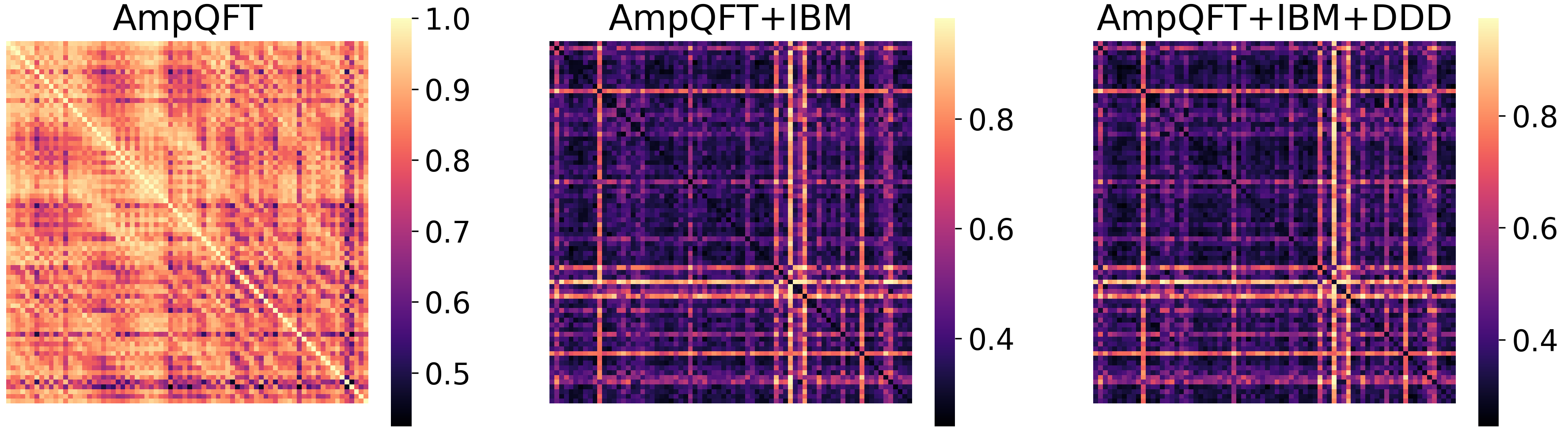}
        \caption{Amplitude embedding + QFT}
        \label{fig:heatmap-qft}
    \end{subfigure}
    \caption{Quantum and classical distance metrics}
    \label{fig:heatmaps}
\end{figure}

We plotted heatmaps of each metric in Fig. \ref{fig:heatmaps}; each classical and quantum metric can be considered a distance.
Correlation has the most variation between neurons, while Fidelity and Euclidean show similar distributions of distances despite having inverted values. For the quantum state fidelities, Ang's checkerboard pattern shows that there are more neuron clusters than the patterns made with classical metrics while also still maintaining overall structure. Amp and Amp+QFT have more granular appearances compared to Euclidean and Fidelity, whereas Ang+IBM, Amp+IBM, and Amp+QFT+IBM appear similar to Euclidean and Fidelity. Applying DDD to Amp and Amp+QFT results in a similar distance matrix as without DDD, both on a simulator and on IBM's QPU.

\subsection{Statistical comparison of distance matrices}

We obtain Spearman rank correlations between each pair of distance matrices with Mantel's test \cite{mantel1967detection} using SciKit-Bio (\href{https://scikit.bio/}{https://scikit.bio/}). Because different metrics may not have the same scale or direction, we plot the absolute value of these correlations in Fig. \ref{fig:mantel-statistic}, where higher values indicate stronger correlation. Key findings include:

\begin{figure}[hb!]
    \centering
    \begin{subfigure}{0.225\textwidth}
        \centering
        \includegraphics[clip,trim={1.5pct 1.5pct 12.5pct 0},width=\textwidth]{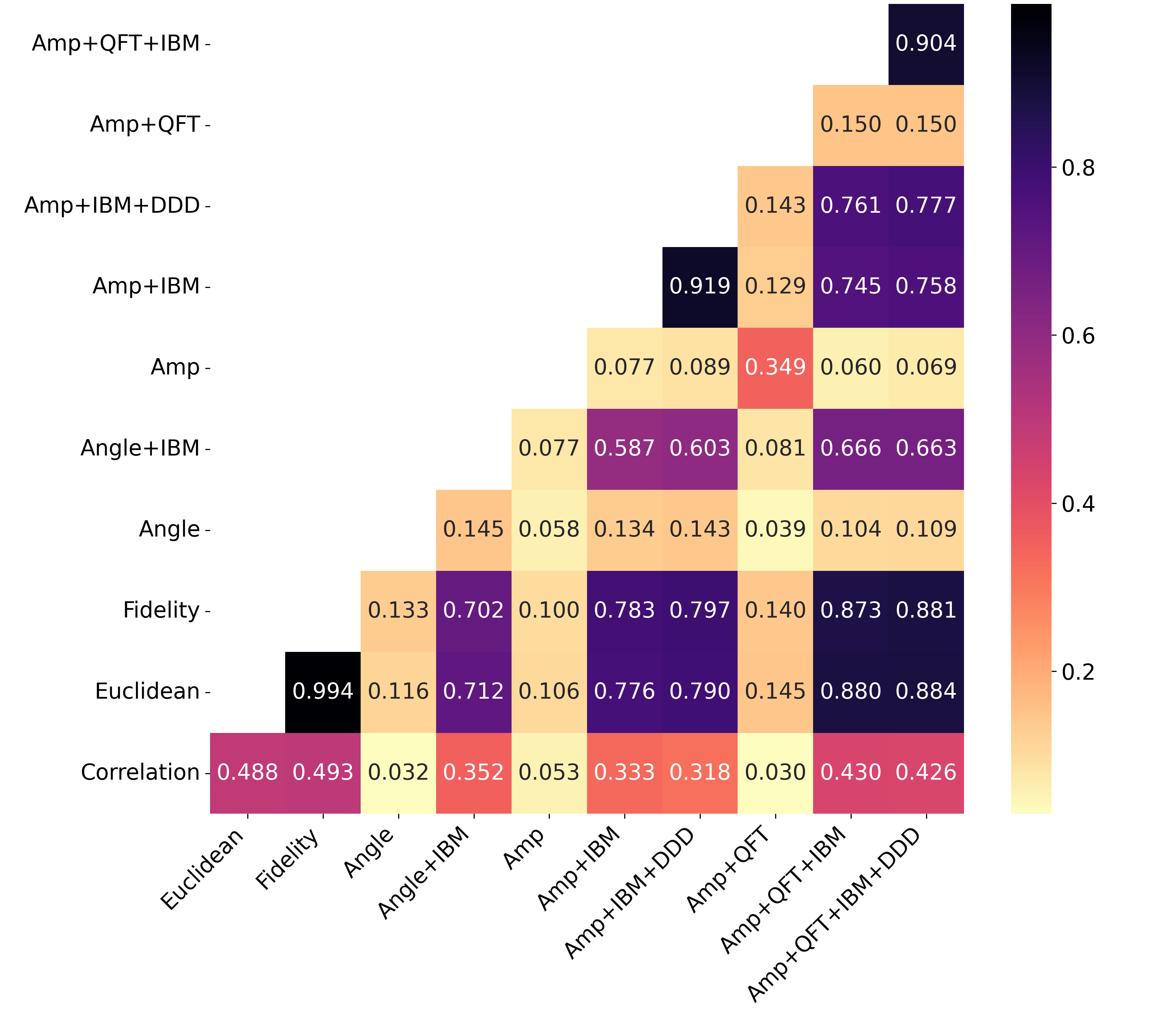}
        \caption{Test statistic}
        \label{fig:mantel-statistic}
    \end{subfigure}
    ~
    \begin{subfigure}{0.225\textwidth}
        \centering
        \includegraphics[clip,trim={1.5pct 1.5pct 12.5pct 0},width=\textwidth]{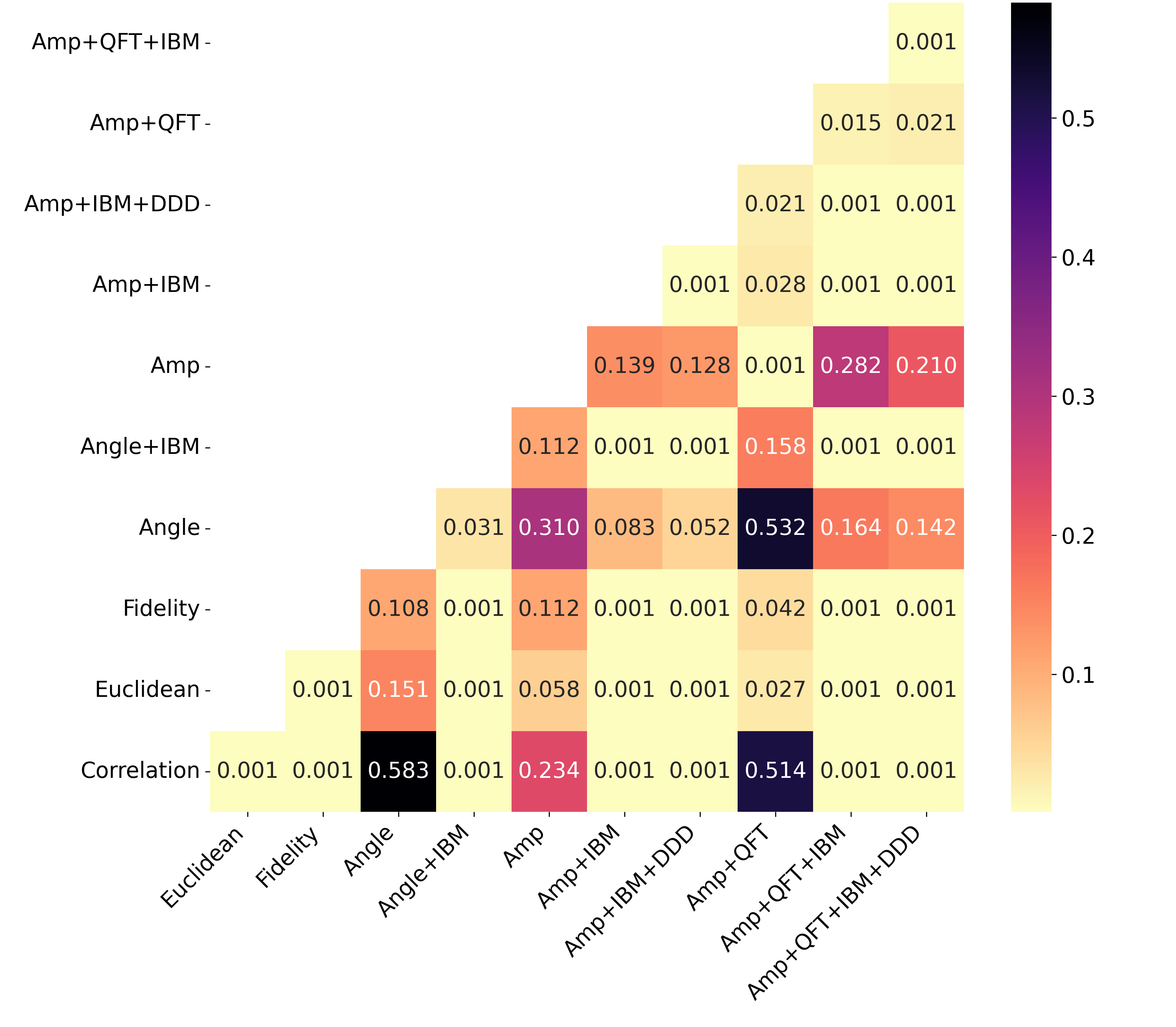}
        \caption{$p$-value}
        \label{fig:mantel-p}
    \end{subfigure}
    \caption{Mantel's test between distance matrices}
    \label{fig:mantel}
\end{figure}

\begin{itemize}
    \item Classical metrics are strongly correlated with each other. The strongest correlation is Euclidean and Fidelity ($r=0.994$, $p=0.001$) and the weakest is Euclidean and Correlation ($r=0.488$, $p=0.001$).
    \item Quantum metrics on Pennylane simulators are moderately correlated with each other, with the strongest correlation between Amp and Amp+QFT ($r=0.349$, $p=0.001$) and the weakest correlation between Amp+QFT and Angle ($r=0.039$, $p=0.532$).
    \item Quantum and classical metrics are weakly correlated with each other, with the strongest correlation being between Euclidean and Amp+QFT ($r=0.145$, $p=0.027$) and the weakest correlation being Correlation and Amp+QFT ($r=0.030$, $p=0.514$)
    \item Quantum metrics on IBM Kingston are strongly correlated with classical metrics, with the strongest correlation being between Euclidean and Amp+QFT+IBM ($r=0.880$, $p=0.001$) and the weakest correlation being Fidelity and Angle+IBM ($r=0.702$, $p=0.001$)
    \item Quantum metrics with DDD error mitigation are highly correlated with the unmitigated version for both Amp+IBM ($r=0.919$, $p=0.001$) and Amp+QFT+IBM ($r=0.904$, $p=0.001$).
\end{itemize}

In summary, quantum metrics on simulators are distinct from classical metrics. Quantum metrics on hardware are more similar to classical metrics, with or without error mitigation.

\subsection{Visual comparison of functional networks}

\begin{figure}[t]
    \centering
    \begin{subfigure}{0.5\textwidth}
        \centering
        \includegraphics[width=0.9\textwidth]{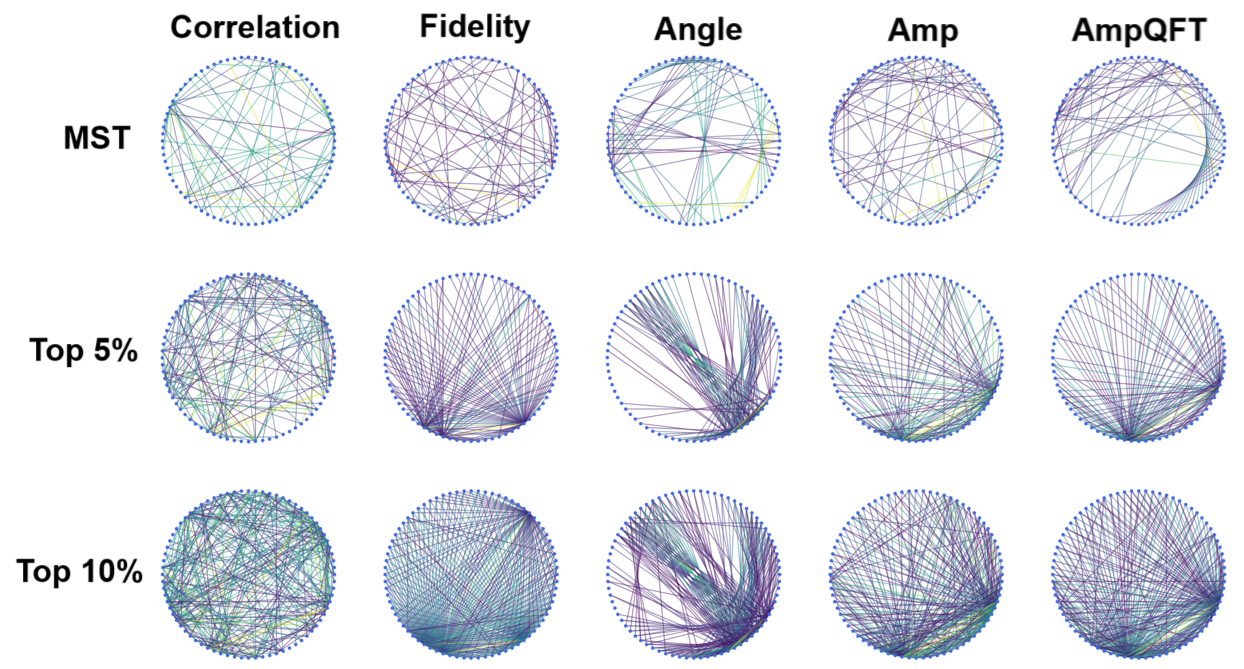}
        \caption{2D networks}
        \label{fig:network-2d}
    \end{subfigure}
    \\
    \begin{subfigure}{0.5\textwidth}
        \centering
        \includegraphics[width=0.9\textwidth]{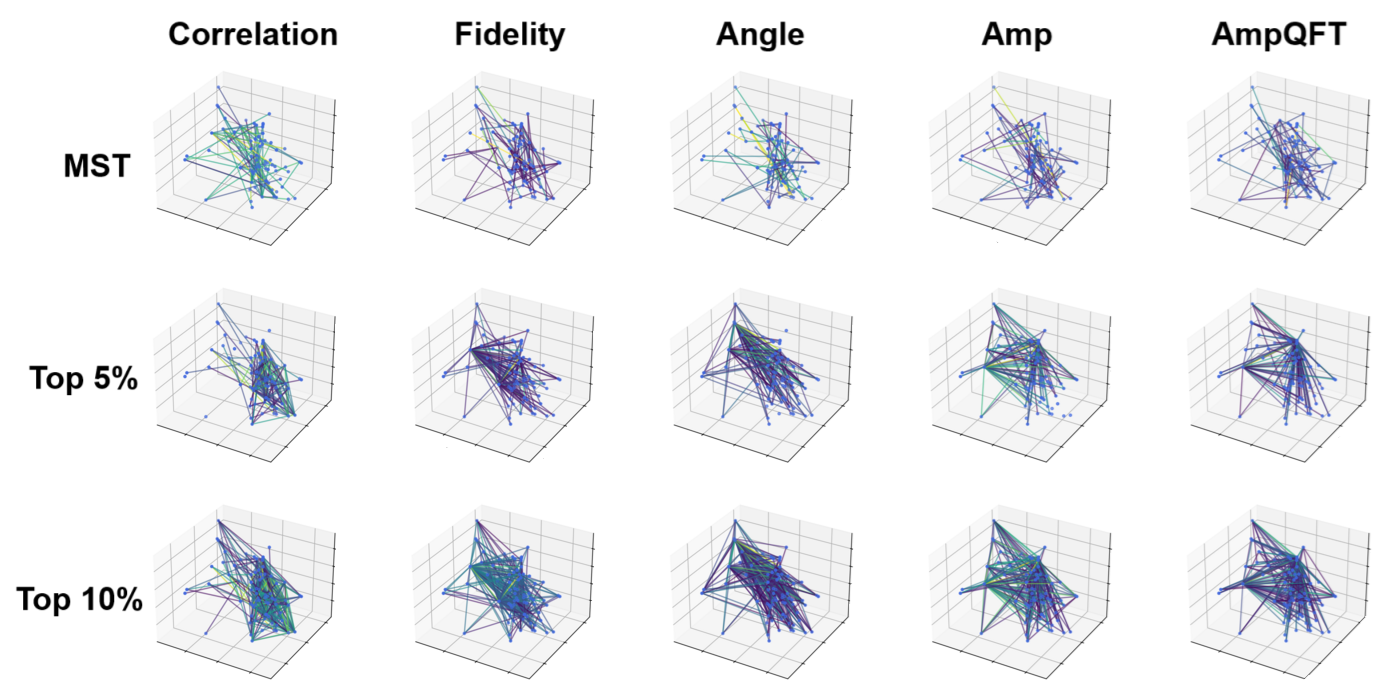}
        \caption{3D networks}
        \label{fig:network-3d}
    \end{subfigure}
    \caption{Functional networks}
    \label{fig:networks}
\end{figure}

For each of Correlation, Fidelity, Ang, Amp, and Amp+QFT, we constructed three kinds of functional networks for each selected metric: the minimum spanning tree (MST) and networks with the top 5\% and 10\% of edges by weight. These networks look different in 2D and 3D space (Fig. \ref{fig:networks}), suggesting that the configurations are potentially different from those of classical techniques.

\section{Discussion}

Identifying functional networks is a necessary step for applications in drug development, brain-computer engineering, and neurotechnology. We applied quantum state fidelities to find functional neural networks that are uncorrelated and difficult to detect with classical metrics like Pearson correlation or Euclidean distance. Our approach highlights the future utility of QC to enhance neuroscience research and applications.

Quantum simulations of Ang, Amp, and Amp+QFT circuits captured major functional connections between neurons while revealing novel deviations from classical metrics. Notably, the identification of a weak correlation between quantum metrics and classical metrics (particularly classical fidelity) highlights how quantum-derived networks have more granular and differentiated network structures. This could inspire future quantum algorithms that complement their classical counterparts with alternative data representations. Our findings also demonstrate our method's capacity to capture non-linear, high-order relationships in neural tuning data. This ability may have significant future applications in neuroscience, as the visually distinct functional networks we constructed may differ in terms of neural activity and relevance.

Our results on the IBM Kingston QPU demonstrate that quantum state fidelity circuits can be run efficiently on current NISQ hardware and point toward future scalability of our method on larger neural connectomes and quantum embeddings. Notably, each state fidelity calculation can be computed computed independently for each pair of neurons, which enables parallel execution across multiple devices and hardware platforms. Our method may also be robust to noise, as distance metrics computed with DDD were not significantly different from those without DDD. However, as quantum metrics were more correlated with classical metrics on hardware than on simulators, further improvements in quantum hardware are needed before achieving clinical relevance.

Limitations of this work include the small number of neurons (76), whereas the full recording contains hundreds of neurons. The tuning curve data are also relatively small with 1 data point per frequency, whereas the raw 2P-imaging waveform can have thousands of data points taken over the course of several minutes. Additionally, with only 1,000 shots on IBM QPUs, the precision of state fidelity estimates could be improved. Finally, real functional connectomes are directed graphs, not undirected. Quantum state fidelities are symmetric and cannot capture this directed nature without modification. Future work will expand analysis to larger neural recordings, use more shots to improve fidelity estimation, and improve representation of directed neuron relationships.

\section{Acknowledgments}

This research is based on the submission by team \href{https://github.com/0mWh/pqic-gic-quadrigems}{Quadrigems (https://github.com/0mWh/pqic-gic-quadrigems)} to the \href{https://web.archive.org/web/20250818014240/https://qlab.umd.edu/research/initiatives/neuroquantum}{NeuroQuantum Nexus} track of the \href{https://web.archive.org/web/20250622115932/https://www.pqic.org/challenge}{Global Industry Challenge 2025} (\href{https://web.archive.org/web/20250622115008/https://gcell.umd.edu/}{gcell.umd.edu}).
The authors thank Ricky Young and Pranet Sharma from qBraid for providing IBM credits, and Maya Ma and Shiraz Robinson II for initial research discussions.
The authors acknowledge the use of generative AI tools for upscaling Fig. \ref{fig:main-figure} and manuscript editing, and take full responsibility for the content of this work.

\bibliographystyle{IEEEtran}
\bibliography{paper}

\end{document}